\documentclass[aps,prb,twocolumn,superscriptaddress,amsmath,showpacs]{revtex4}
\usepackage[pdftex]{graphicx}

\begin{document}
\title{Quasiparticle calculations of the electronic properties of ZrO$_2$ and
HfO$_2$ polymorphs and their interface with Si}
\author{Myrta Gr\"uning} 
\altaffiliation[Present address: ]{Centre for Computational Physics and
  Physics Department, University of Coimbra, Rua Larga 3004-516
  Coimbra (Portugal)}
\affiliation{European Theoretical Spectroscopy Facility (ETSF) and
  Unit\'e PCPM, Universit\'e Catholique de Louvain, Place Croix du Sud
  1, 1348 Louvain-la-Neuve (Belgium)}
\author{Riad Shaltaf}
\altaffiliation[Present address: ]{Department of Physics,
  University of Jordan, 11942, Amman (Jordan)}
\affiliation{European Theoretical Spectroscopy Facility (ETSF) and
  Unit\'e PCPM, Universit\'e Catholique de Louvain, Place Croix du Sud
  1, 1348 Louvain-la-Neuve (Belgium)} 
\author{Gian-Marco Rignanese}
\affiliation{European Theoretical Spectroscopy Facility (ETSF) and
  Unit\'e PCPM, Universit\'e Catholique de Louvain, Place Croix du Sud
  1, 1348 Louvain-la-Neuve (Belgium)}
\affiliation{}
\date{\today}

\begin{abstract}

Quasiparticle calculations are performed to investigate the electronic
band structures of various polymorphs of Hf and Zr oxides. The
corrections with respect to density-functional-theory results are found
to depend only weakly on the crystal structure. Based on
these bulk calculations as well as those for bulk Si, the effect of
quasiparticle corrections is also investigated for the band offsets at
the interface between these oxides and Si assuming that the
lineup of the potential at the interface is reproduced correctly within
density-functional theory. On the one hand, the valence band offsets
are practically unchanged with a correction of a few tenths of eV. On
the other hand, conduction band offsets are raised by 1.3--1.5\,eV.
When applied to existing calculations for the offsets at the
density-functional-theory level, our quasiparticle corrections provide
results in good agreement with the experiment.

\end{abstract}

\pacs{71.10.-w,73.40.Ty,73.40.Lq,85.40.-e,85.60.-q}

\maketitle

\section{Introduction}
In the microelectronic industry, the continuous quest for devices with
improved performance and lower power consumption has recently stimulated an
intense research on dielectric materials. Indeed, for over three decades,
SiO$_2$ has formed the perfect gate dielectric material for metal oxide
semiconductor field effect transistors (MOSFETs). However, fundamental limits
have recently been reached that impede further downscaling of MOSFETs based
on SiO$_2$.~\cite{ Wilk2001} Candidate materials to substitute the latter are
transition metal oxides and silicates with an high dielectric constant,
specifically higher than SiO$_2$, and are commonly referred to as
\emph{high-k} dielectrics. In this framework, ZrO$_2$ and HfO$_2$, and more
generally Zr and Hf compounds, have attracted considerable attention,~\cite{
Wilk2001} hafnium-based microprocessors being now in development or even
already in production.~\cite{ note1}

\emph{Ab-initio} calculations can nicely complement the experimental
work to investigate the properties of these novel materials (see, e.g.,
Refs.~\onlinecite{ Rignanese2001, Rignanese2004, Rignanese2004a}) and to
engineer the interfaces.~\cite{ Peressi1998, Wang2006a} The method of
choice for investigating ground-state properties is density-functional
theory (DFT) that allows to treat quite large systems on the one hand,
and to obtain reliable results on the other hand.~\cite{ Demkov2007} For
the Si/ZrO$_2$ and Si/HfO$_2$ interfaces, various models have been
explored using DFT.~\cite{ Peacock2004, Peacock2006,
Puthenkovilakam2004a, Puthenkovilakam2004b, Fiorentini2002, Dong2005,
Tuttle2007} It was found that due to their analogous electronic
structure, the two transition metal oxides present a very similar interfacial
bonding. Moreover, there is general agreement that the O-terminated
interfaces are more stable than metal-terminated ones.

One of the most stringent criteria in the design of Si/oxide interfaces
is their band offsets (BOs) that control the transport properties, and
hence the leakage current.~\cite{ Wilk2001} In particular, both the
valence and conduction band offsets (VBO and CBO) should be larger than
1\,eV to obtain a low leakage. DFT relying on local or semilocal
approximations for the exchange--correlation potential does not
guarantee quantitatively correct BOs since the DFT eigenenergies do not
correspond to the quasiparticle (QP) energies.~\cite{ PerdewL83,
ShamS83, GruningMR06a, GruningMR06b, note2} However, the VBOs are often found
with an accuracy of a few tenths of eV, especially for semiconductor
interfaces.~\cite{ Van1987} Therefore, the CBOs can also be predicted
using a simple scissor operator to correct the band gaps to their
experimental values. Several works have addressed the BOs at the
Si/ZrO$_2$ and Si/HfO$_2$ interfaces using this scissor-corrected DFT
scheme.\cite{Peacock2004, Peacock2006,
Puthenkovilakam2004a, Puthenkovilakam2004b, Chen2009} The calculated
VBOs for the stable insulating O-terminated interfaces of Si/ZrO$_2$ and
Si/HfO$_2$ are around 2.5--3\,eV, in reasonable agreement with the
experiments (2.7--3.4\,eV).~\cite{ Miyazaki2001, Oshima2003, Wang2004,
Rayner2002, Sayan2002, Afanasev2002, Sayan2004a, Renault2004,Bersch2008}
The scissor-corrected CBOs are about 1.7--2.2\,eV, and compare quite
well with the experimental values (1.5--2\,eV).~\cite{ Afanasev2002,
Sayan2004a, Renault2004, Bersch2008}

In contrast with DFT, the many-body perturbation theory (MBPT) within
the $GW$ approximation has proven to be a practical and
sufficiently accurate method for calculating QP energies.~\cite{
AryasetiawanG98, Hedin99, AulburJW00} In this method, the DFT
eigenenergies within the local density approximation (LDA) or the
generalized gradient approximation (GGA) level are corrected
perturbatively (QP corrections) to obtain the QP energies. For
semiconductor interfaces, the QP corrections on the band edges are often
similar on both sides~\cite{ Zhu1991} and do not substantially affect
the VBOs, explaining the success of DFT.~\cite{ Van1987} Nevertheless,
this approach cannot be generalized to other interfaces. For example,
for the Si/SiO$_2$ interface, the difference between the DFT and the
experimental VBO is larger than 1\,eV.  Recent accurate calculations
have shown that the QP corrections account for this discrepancy,~\cite{
Shaltaf2008} and hence they are essential to reproduce quantitatively
the experimental measurements. For the Si/ZrO$_2$ interface, a
correction of about 1.1\,eV has been extracted from $GW$
calculations for Si~\cite{Zhu1991} and
ZrO$_2$~\cite{Kralik1998} and used together with the experimental band
gap to correct DFT BOs in several works.~\cite{ Fiorentini2002, Dong2005}
For the Si/HfO$_2$ interface, the same correction as for Si/ZrO$_2$ has
been adopted~\cite{ Tuttle2007} since there were no $GW$ calculations
available for HfO$_2$. Such an assumption seems quite reasonable given
the analogous electronic structure of ZrO$_2$ and HfO$_2$. However, for
both Si/ZrO$_2$ and Si/HfO$_2$ interfaces, the VBOs obtained applying
this correction are too large (and as a consequence the CBOs too small)
with respect to the available experiments.~\cite{ Fiorentini2002,
Dong2005, Tuttle2007}

In this work, the origin of this disagreement is discussed. QP calculations
are performed for the thermodynamically stable phases (cubic, tetragonal and
monoclinic) of HfO$_2$ and ZrO$_2$ as well for a strained tetragonal
polymorph. In particular, we calculate the QP corrections at the top valence
and bottom conduction bands in order to determine the fundamental band gaps
and by comparison with Si, the QP correction for the Si/oxide band offsets.
In contrast with previous QP calculations, our results show that the VBO
remains almost unchanged while the CBO is corrected by 1.3--1.5\,eV, thus
explaining the success of the scissor-corrected DFT. The paper is organized
as follows. In Sec. II, the methodological background is briefly described.
Sec. III is devoted to the presentation and discussion of our results: the
DFT geometries and band structures, the QP corrections to the band gaps, and
finally the QP corrections to the Si/oxide band offsets.

\section{Methodological background}
\label{sc:method}

The geometries and electronic structures for all the systems are computed
within the DFT.  All the calculations are carried out with the
\textsc{abinit}~\cite{ abinitref} code within the LDA for the
exchange--correlation energy functional.~\cite{ CeperleyA80}
Troullier-Martins~\cite{ TroullierM91} norm conserving pseudopotentials are
used which include semicore states in the pseudopotentials for Zr and Hf (for
details see Refs.~\onlinecite{ Rignanese2001, Rignanese2004}). The wave
functions are expanded on a plane-wave basis set up to kinetic energy cutoff
of 12 Ha for Si and up to 30 Ha for the HfO$_2$ and ZrO$_2$ polymorphs.  For
all systems, the Brillouin zone (BZ) is sampled with a 4$\times$4$\times$4
Monkhorst--Pack~\cite{ MonkhorstP76} grid.

The QP energies are evaluated using the MBPT within the $GW$
approximation. In this approach, the DFT eigenenergy
$E_n^{\text{DFT}}$ and wavefunction $\psi_n^{\text{DFT}}$ for the
$n^\text{th}$ state are used as a zeroth-order approximation for their
quasiparticle counterparts. Thus, the QP energy $E_n^{\text{QP}}$  is
calculated  by adding to $E_n^{\text{DFT}}$ the first-oder
perturbation correction that comes from replacing the DFT
exchange-correlation potential $v_{\text{xc}}^{\text{DFT}}$ with the
$GW$ self-energy operator $\Sigma_{GW}$:
\begin{equation}
   E_n^{\text{QP}} = E_n^{\text{DFT}} +
  \Re\{ Z_n  \langle
   \psi_n^{\text{DFT}}| 
   \Sigma_{GW} - v_{\text{xc}}^{\text{DFT}}
   |\psi_n^{\text{DFT}} \rangle\}.
\label{eq:qpcrrct}
\end{equation}
The renormalization factor $Z_n$ accounts for the fact that
$\Sigma_{GW}$, which is energy dependent, should be evaluated at
$E^{\text{QP}}_n$. The $GW$ self-energy operator $\Sigma_{GW}$ writes
as a convolution in frequency space between the one-electron
Green's function $G$ and the screened Coulomb potential $W$:
\begin{equation}
\Sigma_{GW} (\textbf{r},\textbf{r}'; \omega)  =
\frac{i}{2\pi} \int d\omega' e^{i\delta \omega'}  G(\textbf{r},\textbf{r}';
\omega') W(\textbf{r},\textbf{r}'; \omega - \omega'),
\end{equation}
where $\delta$ is a positive infinitesimal.
The explicit expression for the Green's function $G$ is:
\begin{equation}
G(\textbf{r},\textbf{r}'; \omega) = \sum_n
\frac{
 \psi_n^{\text{DFT}}(\textbf{r})
 \left[ \psi_n^{\text{DFT}}(\textbf{r}') \right]^*
}{
 \omega -E_n^{\text{DFT}} + i\delta\,\text{sgn}(E_n^{\text{DFT}} - \mu)
},
\label{eq:greens}
\end{equation}
where $\mu$ is the
chemical potential.
The screened Coulomb potential is determined as convolution between the
inverse of the dielectric function $\epsilon^{-1}$ and the bare Coulomb
interaction:
\begin{equation}
W(\textbf{r},\textbf{r}'; \omega)  = \int d\textbf{r}'' \frac{\epsilon^{-1}
(\textbf{r},\textbf{r}''; \omega)}{|\textbf{r}'' - \textbf{r}'|}. 
\label{eq:scrclb}
\end{equation}
The dielectric function $\epsilon$ is calculated within the Random Phase
Approximation. Its dependence on the frequency is approximated using the
plasmon pole model (PPM) proposed by Godby and Needs.~\cite{ Godby1989} For
the cubic ZrO$_2$, we explicitly test the validity of this choice. On the one
hand, we perform the same calculation with the PPM proposed by Hybertsen and
Louie.~\cite{ HybertsenL86} On the other hand, we take into account the full
frequency dependence of the dielectric matrix without resorting to any PPM at
all,~\cite{ MariniODS02, ThesisMarini} in order to discriminate between the
two PPMs. 

In our QP calculations, both $G$ [Eq.~(\ref{eq:greens})] and $W$
[Eq.~(\ref{eq:scrclb})] are first evaluated from the DFT eigensolutions
(which is often referred to as $G_0W_0$). Successively, the DFT energies are
self-consistently replaced in Eq.~(\ref{eq:greens}) by the corrected values
obtained from Eq.~(\ref{eq:qpcrrct}) ($GW_0$). The QP corrections are
evaluated only for a few valence and conduction bands around the gap, the
other energies $E_n^{\text{QP}}$ are extrapolated using a scissor operator.
We find that 2--3 iterations are enough to converge the orbital energies up
to 0.01--0.02\,eV. A systematic study~\cite{ Shishkin2007} on several bulk
systems has recently pointed out $GW_0$ as a practical and accurate method
for evaluating QP energies.  Indeed, while $G_0W_0$ provides underestimated
energies for almost all systems, $GW_0$ shows a good agreement with both
experimental data and full self-consistent $GW$ including a
Bethe--Salpeter-like vertex correction.~\cite{ Shishkin2007a}

The $GW$ calculations are performed with the \textsc{yambo}~\cite{ yamboref}
code, except for the calculation of cubic ZrO$_2$ with the PPM of Hybertsen
and Louie,~\cite{ HybertsenL86} which is performed with the \textsc{abinit}
code. We carefully study the convergence of the QP corrections with the
numerical cutoffs: for Si and the cubic polymorphs of HfO$_2$ and ZrO$_2$, we
include 200 bands in the calculations of the Green's function
[Eq.~(\ref{eq:greens})], and 200 bands and 331 reciprocal lattice vectors in
the calculation of the dielectric matrix~\cite{ note3} in
Eq.~(\ref{eq:scrclb}). For the tetragonal polymorphs (strained and at
equilibrium), we include 500 bands in the calculations of the Green's
function, and 300 bands and 735 reciprocal lattice vectors in the calculation
of the dielectric matrix. Finally, the monoclinic polymorphs requires 600
bands for the Green's function, and 400 bands and 1177 reciprocal lattice
vectors for the dielectric matrix. With these parameters, we estimate an
error on the QP energies of about 0.05--0.1\,eV depending on the system. 

\section{Results}
\subsection{DFT geometries and band structures}

For both ZrO$_2$ and HfO$_2$, the three thermodynamically stable phases
(cubic [c], tetragonal [t], and monoclinic [m]) are investigated.  In
addition, a strained [s] form of the tetragonal polymorph is also considered
in which two sides are fixed to $a=a_{\text{Si}}/\sqrt{2}$
($a_{\text{Si}}$=5.40 \AA\ is the LDA theoretical lattice constant of Si)
while all other degrees of freedom are relaxed. This last structure aims at
simulating the effect of the epitaxial strain on the oxide in the MOSFET
device.

The calculated equilibrium parameters describing these four geometries are
reported in Table~\ref{tb:geores}. Our results agree within 1--2\% with
previous LDA~\cite{ Kralik1998, Dash2004, Puthenkovilakam2004a,
Puthenkovilakam2004b} and GGA~\cite{ Foster2001, Jaffe2005, Dong2005,
Terki2005, Mukhopadhyay2006, Garcia2006, Tuttle2007} results, as well as with
experimental data.~\cite{ note4} For the strained polymorphs (s-ZrO$_2$ and
s-HfO$_2$), a contraction is observed along the tetragonal direction $c$ of
about 2\% compared to their fully relaxed tetragonal analogs (t-ZrO$_2$ and
t-HfO$_2$). This is a direct consequence of fixing the lattice constant $a$ in
the basal plane (in fact, $a$ is expanded by 7\% and 5\% in s-ZrO$_2$ and
s-HfO$_2$, respectively). As an indirect consequence, the internal parameter
$d_z$ becomes larger in the strained forms.  For the latter structure, no
direct meaningful comparison is possible with previous works in which
epitaxial strained tetragonal polymorphs are also considered, ~\cite{
Dong2005, Tuttle2007} since these rely on the GGA value for Si lattice
constant. 

\begin{table}[h]
\begin{ruledtabular}
\begin{tabular}{clllrrllrrl}
  & &                  & & \multicolumn{3}{c}{ZrO$_2$}
                       & & \multicolumn{3}{c}{HfO$_2$}\\
\cline{5-7} \cline{9-11}
c & & $a$              & & 5.011 &       &
                       & & 5.273 &       &\\
t & & $a$ \ $c$ \ $d_z$& & 3.547 & 5.086 & 0.040
                       & & 3.616 & 5.169 & 0.031 \\
s & & $a$ \ $c$ \ $d_z$& & 3.817 & 4.980 & 0.058
                       & & 3.817 & 5.053 & 0.041 \\
m & & $a$ \ $b$ \ $c$  & & 5.050 & 5.185 & 5.190
                       & & 5.171 & 5.276 & 5.292 \\
  & & $\gamma$         & & 99.09$^\circ$ & &
                       & & 99.27$^\circ$ \\
  & & $M$              & & \multicolumn{3}{l}{(0.2780 0.0416 0.2097)}
                       & & \multicolumn{3}{l}{(0.2778 0.0404 0.2059)}\\
  & & O$_1$            & & \multicolumn{3}{l}{(0.0789 0.3527 0.3279)}
                       & & \multicolumn{3}{l}{(0.0799 0.3527 0.3277)}\\
  & & O$_2$            & & \multicolumn{3}{l}{(0.4460 0.7594 0.4838)}
                       & & \multicolumn{3}{l}{(0.4462 0.7600 0.4857)}\\
\end{tabular}
\end{ruledtabular}
\caption{Structural parameters of the thermodynamically stable phases (cubic
[c], tetragonal [t], and monoclinic [m]) and of the strained [s] tetragonal
structure (see text) of ZrO$_2$ and HfO$_2$.  The lattice constants ($a$,
$b$, and $c$) are in expressed in \AA, while the angle $\gamma$ (between $a$
and $b$) is given in degrees.  For the tetragonal and strained forms, the
internal parameter $d_z$ is the displacement of the oxygen atoms with respect
to their ideal cubic position in units of the lattice vector $c$. For the
monoclinic polymorph, the internal coordinates for the metal ($M$=Zr or Hf)
and the two oxygen (O$_1$ and O$_2$) atoms are given in terms of lattice
vectors.}
\label{tb:geores}
\end{table}

\begin{figure*}
\includegraphics{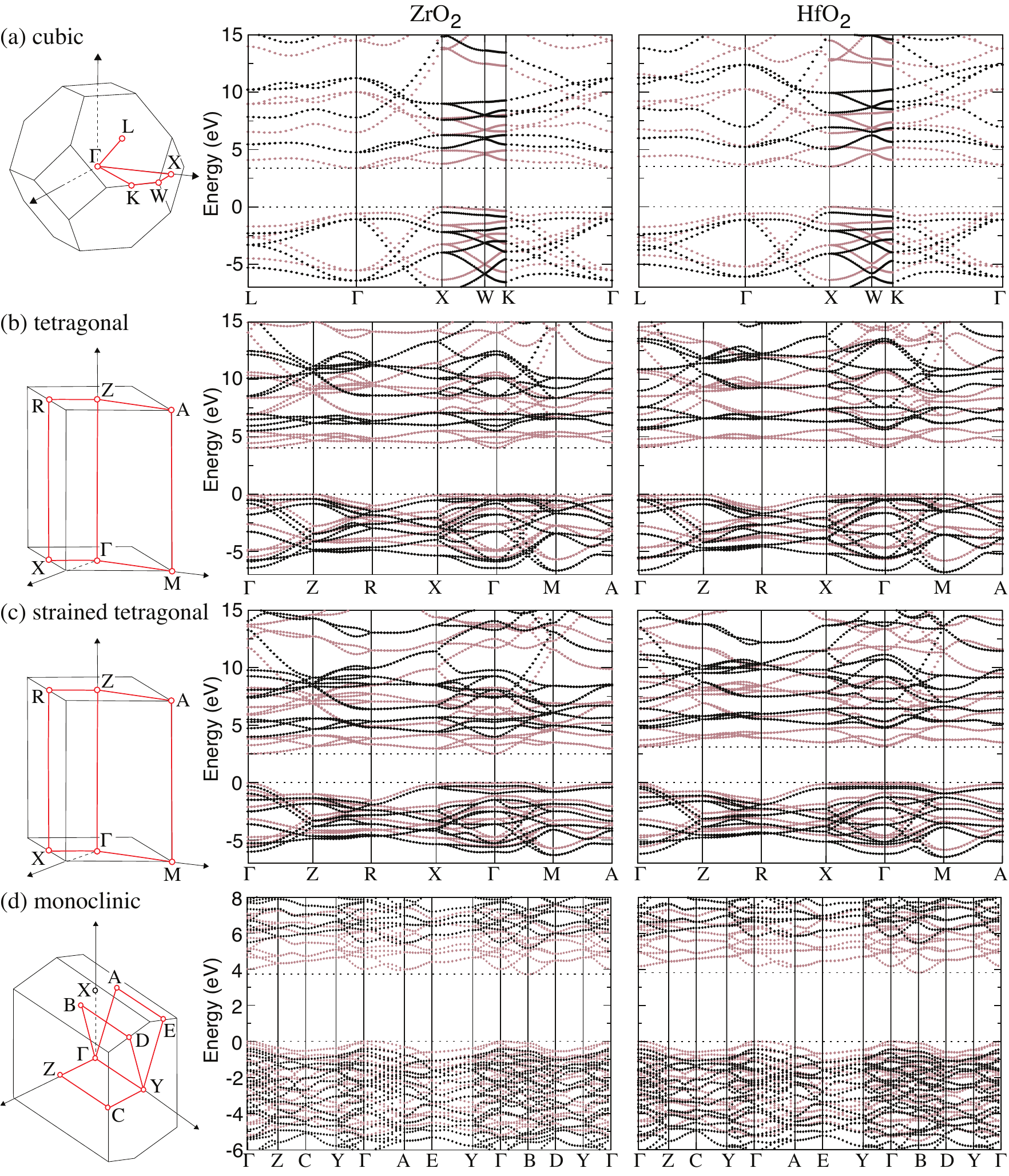}
\caption{[color online] Theoretical band structure within the LDA (brown
circles) and $GW_0$ (black circles) along high symmetry axis of the BZ for
the (a) cubic, (b) tetragonal, (c) strained, and (d) monoclinic phases of
ZrO$_2$ and HfO$_2$. The $GW_0$ band structures have been extrapolated from
the calculated QP corrections using a linear fit.}
\label{fig:bndds} 
\end{figure*}

The band structure calculated within the LDA are reported in brown in
Fig.~\ref{fig:bndds} for the cubic, tetragonal, strained, and monoclinic
forms of ZrO$_2$ and HfO$_2$.  For ZrO$_2$, the LDA band structures are in
good agreement with those presented in Ref.~\onlinecite{ Dash2004}.  The band
structures of HfO$_2$ are overall very similar to those of ZrO$_2$, with the
important exception of the cubic phase [Fig.~\ref{fig:bndds}(a)].  Indeed,
while c-ZrO$_2$ shows an indirect minimum band gap from X (top of the
valence band) to $\Gamma$ (bottom of the conduction band) of 3.4\,eV,
c-HfO$_2$ has a direct band gap at X of 3.5\,eV. The indirect X-$\Gamma$
gap ---that is the minimum gap in c-ZrO$_2$--- is slightly larger (3.7\,eV).

In the tetragonal phase [Fig.~\ref{fig:bndds}(b)], the top valence band is
almost flat along the $\Gamma$Z and X$\Gamma$M directions. It actually
presents several maxima (close to X at $\sim$0.20 in the X$\Gamma$
direction and close to $\Gamma$ at $\sim$0.20 in the $\Gamma$M direction,
in A and Z). The conduction band minimum is located at $\Gamma$. The
resulting indirect band gap is 4\,eV for both t-ZrO$_2$ and t-HfO$_2$,
respectively.  The effect of the strain is to reduce the band gap to 2.5\,eV
for ZrO$_2$, and to 3.1\,eV for HfO$_2$ [Fig.~\ref{fig:bndds}(c)]. The effect
is larger for s-ZrO$_2$ since it has the larger mismatch with the Si lattice
constant.  A similar reduction of the band gap resulting from the Si-epitaxial
strain was also observed in previous works.~\cite{ Dong2005, Tuttle2007} For
both s-ZrO$_2$ and s-HfO$_2$, the valence band is almost flat along the
X$\Gamma$M direction. The maxima are located close to $\Gamma$, while the
conduction band minimum is at $\Gamma$.  

For the monoclinic phase [Fig.~\ref{fig:bndds}(d)], the top of the valence
band is located at $\Gamma$ while the bottom of the conduction band is
located at B. The indirect minimum band gap is 3.7\,eV and 3.8\,eV for
m-ZrO$_2$ and m-HfO$_2$, respectively.

\subsection{QP corrections on the band gaps}

For all systems, the QP corrections are reported in Table~\ref{tb:allmqp} and
the resulting band structures appear in black in Fig.~\ref{fig:bndds}. The
effect of the QP corrections is to lower the valence bands and to raise the
conduction bands. For both ZrO$_2$ and HfO$_2$, the QP corrections on the
band edges depend only weakly on the structure (c, t, s, or m). The
correction $\delta E_v$ at the valence band maximum (VBM) varies from $-$0.3
to $-$0.5\,eV in ZrO$_2$ and from $-$0.4 to $-$0.5\,eV in HfO$_2$, while the
correction $\delta E_c$ at the conduction band minimum (CBM) ranges from 1.4
to 1.5\,eV in ZrO$_2$ and from 1.5 to 1.7\,eV in HfO$_2$.  Moreover, the QP
corrections are not affecting the location of the minima and maxima found in
the LDA bandstructures. The net effect of the QP corrections is thus to open
the LDA band gaps by $\delta E_g$=$\delta E_c-\delta E_v$ varying from 1.8 to
2\,eV for ZrO$_2$ and from 1.9 to 2.1\,eV for HfO$_2$. The difference between
the ZrO$_2$ and HfO$_2$ band gaps found in LDA is increased by $\sim$0.2\,eV
when including the QP corrections (from 0.1 to 0.3\,eV in the cubic and
monoclinic phase, from 0.6 to 0.8\,eV in the strained phase), except for the
tetragonal phase in which the band gaps differ by less than 0.1\,eV both at
the LDA and QP level.

\begin{table}[h]
\begin{ruledtabular}
\begin{tabular}{lrlrrrrlrrrr}
 & \multicolumn{1}{c}{Si} &
 & \multicolumn{4}{c}{ZrO$_2$} &
 & \multicolumn{4}{c}{HfO$_2$} \\
\cline{4-7} \cline{9-12}
 & &    
 & \multicolumn{1}{c}{c} & \multicolumn{1}{c}{t}
 & \multicolumn{1}{c}{s} & \multicolumn{1}{c}{m} &
 & \multicolumn{1}{c}{c} & \multicolumn{1}{c}{t} 
 & \multicolumn{1}{c}{s} & \multicolumn{1}{c}{m}  \\
$E_g^\text{DFT}$
 &   0.4 & &   3.4 &   4.0 &   2.5 &   3.7 & &   3.5 &   4.1 &   3.1 &   3.8 \\
$\delta E_v$
 &$-$0.6 & &$-$0.5 &$-$0.4 &$-$0.3 &$-$0.4 & &$-$0.5 &$-$0.4 &$-$0.4 &$-$0.4 \\
$\delta E_c$
 &   0.1 & &   1.4 &   1.5 &   1.5 &   1.5 & &   1.5 &   1.5 &   1.6 &   1.7 \\
$E_g^\text{QP}$
 &   1.1 & &   5.3 &   5.9 &   4.3 &   5.6 & &   5.5 &   6.0 &   5.1 &   5.9 \\
\end{tabular}
\end{ruledtabular}
\caption{For bulk Si, and bulk ZrO$_2$ and HfO$_2$ in the cubic (c),
tetragonal (t), monoclinic (m) phases and strained (s) polymorph: the minimum
DFT band gap $E_g^{\text{DFT}}$, the minimum QP band gap $ E_g^{\text{QP}}$
at $GW_0$ level and the QP corrections $\delta E_v$ and $\delta E_c$ to the
VBM and CBM, respectively.}
\label{tb:allmqp}
\end{table}

For ZrO$_2$, the QP correction to the band gap, which is ~1.9\,eV in the
three thermodynamically stable phase, is 0.4\,eV lower than the value
previously obtained in Ref.~\onlinecite{ Kralik1998} for the cubic phase.
This variation can be attributed to the different PPM (by Hybertsen
Louie~\cite{ HybertsenL86}) used by the authors in the frequency integration
of Eq.~(\ref{eq:scrclb}). Indeed, by repeating our calculation for the cubic
phase with the same PPM, the QP correction to the band gap increases up to
2.4\,eV in good agreement with Ref.~\onlinecite{ Kralik1998}. In order to
discriminate between the two PPMs, the calculation is also repeated without
resorting to any PPM. The QP correction is found to be 2.1\,eV, which is
0.2\,eV higher than the value obtained with the Godby and Needs PPM, and
0.3\,eV lower than the one obtained with the Hybertsen and Louie PPM. It can
be argued that our QP corrections to the band gap and hence the resulting QP
band gaps are probably also underestimated for the other polymorphs of
ZrO$_2$ and for HfO$_2$ due to the use of the PPM. Thus, when comparing the
results with the experiments or other theoretical work, this extra
uncertainty of about 0.2\,eV should also be taken into account.

Our calculated QP band gaps for c-, t-, and m-ZrO$_2$, also show an overall
agreement with those of Ref.~\onlinecite{ Medvedeva2007} relying on the
screened-exchange LDA method. This confirms the validity of this
approximation for the calculation of band structures. For HfO$_2$, our result
for the cubic phase agrees well with the $GW$ band gap reported in
Ref.~\onlinecite{ Nishitani2007}, while for the tetragonal and monoclinic
phase our values are larger by 0.2\,eV.

Experimentally, the optical band gaps determined from energy loss and
transmission spectroscopies range from 5.2 to 5.7\,eV for ZrO$_2$~\cite{
French1994, Dash2004, Aita2003, Nohira2002, Kosacki1999, Sayan2005,
Bersch2008} and from 5.3 to 5.8\,eV for HfO$_2$.~\cite{ Bersch2008,
Cheynet2007, Perevalov2007, Yu2002, Balog1977} Hence, the agreement between
our calculations of the fundamental gap and experimental results is quite
reasonable considering the temperature, excitonic, and impurities effects and
the possible substrate strain in case of deposited films that should be taken
into account. For the thermodynamically stable phases of ZrO$_2$, reflectance
vacuum ultraviolet spectroscopy indicate 6.1--7.1\,eV, 5.8--6.6\,eV, and
5.8--7.1\,eV for the cubic, tetragonal, and monoclinic phases, respectively.
For the last two, our calculated values fall into the range of the
experimental estimates. In contrast, for the cubic phase, our band gap is
much lower even when considering the minimum direct band gap (5.5\,eV at X)
and when taking into account the underestimation by 0.2\,eV coming from the
PPM. This discrepancy may be related to the yttrium used to stabilize the
cubic phase at room temperature and/or the tendency of reflectance
measurements to overestimate the gap (see Ref.~\onlinecite{ Dash2004} for a
thorough discussion). For HfO$_2$, a ``theoretical'' band gap of
6.7\,eV~\cite{ Sayan2004} was proposed for a film deposited on a
SiO$_x$N$_y$/$p$-Si substrate (monoclinic phase) by comparing direct/inverse
photoemission spectroscopy with DFT density-of-states calculations, arguing
that the reduction of 0.9\,eV with respect to the experimental band gap
(5.9\,eV) should be attributed to defects tail states. Our value compares
better with the band gap determined directly from the experiment.

\subsection{QP corrections on the Si/oxide band offsets} 

In the DFT approach, the VBO and CBO are conveniently split into two terms:
\begin{eqnarray}
\text{VBO}=\Delta E_v^\text{DFT} + \Delta V
\label{eq:vbo_dft}\\
\text{CBO}=\Delta E_c^\text{DFT} + \Delta V
\label{eq:cbo_dft}
\end{eqnarray}
The first term $\Delta E_{v}^\text{DFT}$ [resp. $\Delta E_{c}^\text{DFT}$] on
the right-hand side of Eq.~(\ref{eq:vbo_dft}) [resp.  Eq.~(\ref{eq:cbo_dft})]
is referred to as the {\it band-structure contribution}. It is defined as the
difference between the VBM [resp. the CBM] {\it relative to the average of
the electrostatic potential in each material}.  These are obtained from two
independent standard bulk calculations on the two interface materials.  The
second term $\Delta V$, called the {\it lineup of the average of the
electrostatic potential} across the interface, accounts for all the intrinsic
interface effects. It is determined from a supercell calculation with a model
interface.

Despite the DFT limitations in finding accurate eigenenergies, the VBOs are
often obtained with a very good precision, in particular for
semiconductors.~\cite{ Van1987} This has opened an indirect route to compute
the CBOs through the experimental band gaps using:
\begin{equation}
\text{CBO}=\Delta E_g^\text{exp}+\text{VBO}
\label{eq:cbo_var}
\end{equation}
where $\Delta E_g^\text{exp}$ is the difference between the
experimental values of the band gap of each material. Note that this
equation is equivalent to applying a scissor correction to the
conduction bands on both sides of the interface, as can be seen by
inserting Eq.~(\ref{eq:vbo_dft}) in Eq.~(\ref{eq:cbo_var}):
\begin{equation}
\text{CBO}=\Delta E_c^\text{DFT} + \Delta V
+(\Delta E_g^\text{exp} - \Delta E_g^\text{DFT})
\label{eq:cbo_scissor}
\end{equation}
and comparing with Eq.~(\ref{eq:cbo_dft}).

As discussed in the introduction, this scissor-corrected DFT scheme has
been used in several studies of the Si/ZrO$_2$ and Si/HfO$_2$.~\cite{
Peacock2004, Peacock2006, Puthenkovilakam2004a, Puthenkovilakam2004b}
For the stable insulating O-terminated interfaces of Si/ZrO$_2$ and
Si/HfO$_2$, the VBOs calculated are found to range from 2.5 to 3\,eV, in
good agreement with the experimental findings (2.7--3.4\,eV).~\cite{
Miyazaki2001, Oshima2003, Wang2004, Rayner2002, Sayan2002, Afanasev2002,
Sayan2004a, Renault2004,Bersch2008} Adopting $\Delta
E_g^\text{exp}$=4.7\,eV ($E_g^\text{exp}$= 1.1\,eV for Si and
5.8\,eV for ZrO$_2$ and HfO$_2$) in Eq.~(\ref{eq:cbo_scissor}), the
scissor-corrected CBOs lie thus between 1.7 and 2.2\,eV. This is also in
good agreement with experiments (1.5--2.0\,eV).~\cite{ Afanasev2002,
Sayan2004a, Renault2004, Bersch2008}

In the QP framework, it was often assumed~\cite{ Zhu1991} and it has
recently been proven~\cite{ Shaltaf2008} that the lineup of the
potential $\Delta V$ is already well-described within DFT.  So that,
only the band-structure contribution is modified:
\begin{eqnarray}
\text{VBO}=\Delta E_v^\text{QP} + \Delta V =
\Delta E_v^\text{DFT}+\Delta(\delta E_v) + \Delta V
\label{eq:vbo_qp}\\
\text{CBO}=\Delta E_c^\text{QP} + \Delta V =
\Delta E_c^\text{DFT}+\Delta(\delta E_c) + \Delta V
\label{eq:cbo_qp}
\end{eqnarray}
where $\delta E_v = E_v^\text{QP}
- E_v^\text{DFT}$ (resp. $\delta E_c = E_c^\text{QP} - E_c^\text{DFT}$)
is the quasiparticle correction at the VBM
(resp. CBM) and $\Delta(\delta E_v)$ [resp.
$\Delta(\delta E_c)$] is the corresponding difference between the two
materials.

For the Si/ZrO$_2$ and Si/HfO$_2$ interfaces, no specific $GW$ study
exists as such. However, Fiorentini \emph{et~al.}~\cite{ Fiorentini2002}
evaluated the QP effects on the VBO of Si/ZrO$_2$ combining the value for
tetragonal ZrO$_2$ ($\delta E_v$=$-$1.23\,eV) from Ref.~\onlinecite{
Kralik1998} with the value for Si from Ref.~\onlinecite{ Zhu1991} ($\delta
E_v$=$-$0.15\,eV.  This results in a total correction of $\Delta(\delta
E_v)$=1.08\,eV on the VBO in Eq.~(\ref{eq:vbo_qp}). This value has been used
in several other works,~\cite{ Dong2005, Tuttle2007} even for Si/HfO$_2$
interfaces. For both Si/ZrO$_2$ and Si/HfO$_2$ interfaces, the VBOs obtained
in this way were found to be too large (and as a consequence the CBOs too
small) with respect to the experimental values.~\cite{ Fiorentini2002,
Dong2005, Tuttle2007}

For the oxides, our QP corrections to the DFT valence band $\delta E_v$ vary
from $-$0.3 to $-$0.5\,eV; while for the conductions bands, $\delta E_c$ ranges
from 1.4 to 1.7\,eV. For Si, our QP corrections, which are reported in
Table~\ref{tb:allmqp}, lead to a band gap that agrees well with previous
theoretical works (e.g. see Ref.~\onlinecite{ Shishkin2007}), and with the
experimental value.~\cite{ OrtegaH93} The total QP correction to the gap
$\delta E_g$ of about 0.7\,eV, comes mostly from the downshift of valence
band state ($\delta E_v$=$-$0.6\,eV).  This $\delta E_v$ value is almost the
same as that found for the polymorphs of both ZrO$_2$ and HfO$_2$.
Therefore, the QP correction on the VBOs are only of 0.1-0.2\,eV and the
correction on the CBOs is about 1.3--1.5\,eV. This explains why previous
studies based on scissor-corrected DFT were in such a good agreement with
experimental results.

Turning to previous works~\cite{ Fiorentini2002, Dong2005,
Tuttle2007} that accounted for QP corrections to
the BOs of Si/ZrO$_2$ and Si/HfO$_2$
interfaces using values from prior $GW$ calculations,~\cite{
Zhu1991, Kralik1998} their disagreement with experiments can be
explained as follows. On the one hand, the QP corrections in the oxide
and Si are not consistent since a different approximation has been used
in Eq.~(\ref{eq:scrclb}) for the dielectric function $\epsilon$. The
calculations for Si use a model dielectric function,~\cite{ Zhu1991}
while the calculations for ZrO$_2$ use the Random Phase Approximation.
In particular, the value obtained for Si $\delta E_v$=$-$0.2\,eV is
lower in absolute value with respect to the value found from the Random
Phase Approximation ($\delta E_v$=$-$0.6\,eV), and hence this
inconsistency artificially increases the QP correction on the VBO. On
the other hand, our QP results (in particular, those for c- and
t-ZrO$_2$) differ from those obtained in Ref.~\onlinecite{ Kralik1998}
where it was found that the total QP correction $\delta E_g$=2.3\,eV
resulted from lowering the valence bands by about 1.3\,eV, and raising
the conduction band by about 1.1\,eV. The difference with our results is
due again to the different PPM used. Indeed, repeating the calculations
with the Hybertsen and Louie PPM for the cubic phase, we found a
correction $\delta E_v$ of $-$1.1\,eV for the valence and $\delta E_c$
of $+$1.3\,eV for the conduction in agreement with Ref.~\onlinecite{
Kralik1998}. When performing the same calculation without resorting to
any PPM, we obtain $\delta E_v$=$-$0.7\,eV and $\delta E_c$=1.4\,eV,
meaning that $\delta E_v$ is 0.2\,eV too low for the Godby and Needs
PPM, and 0.5\,eV too high for the Hybertsen and Louie PPM. For the
conduction bands, $\delta E_c$ differs by less than 0.1\,eV with the
former, while it is 0.3\,eV too low with the latter.

As final remark, we stress that rigorously, the QP corrections on the band
offsets should be calculated using the same pseudopotential and the same
exchange-correlation approximation as for the interface calculations. Indeed,
while the $GW$ band gap has been demonstrated to be quite insensitive to the
starting point, this is not true for the QP corrections that reflect---as a
consequence---the differences in the choice of the pseudopotential and the
exchange-correlation approximation. 
Therefore, we would recommend to calculate---when
  possible---QP corrections on DFT band offsets using the same pseudopotentials,
and exchange-correlation approximation as for the interface calculation
and using the same PPM for both materials.

\section{Summary}

The electronic properties of ZrO$_2$ and HfO$_2$ polymorphs and their
interface with Si have been investigated using $GW$ calculations.  The
QP corrections are found to be very similar for the two oxides
consistently with their analogous band structure, and depend only
slightly on the crystalline structure.  While, at the DFT level, the
epitaxial strain was found to dramatically shrink the band gap
(especially for ZrO$_2$ for which the lattice parameter mismatch with Si
is larger), the QP corrections depend only slightly on the strain.  When
considering the interface Si/oxide, the QP corrections to the VBOs were
calculated to be very small (a few tenths of eV) by cancellation of the
corrections on the valence band maximum of the Si and those of oxides.
On the other hand, the correction was found to be of the order of
1.5\,eV for the CBOs. These results disagree with the correction on the
VBOs of more than 1\,eV used in the
literature,\cite{ Fiorentini2002,
Dong2005, Tuttle2007} which was extracted from existing $GW$
calculations.~\cite{ Kralik1998, Zhu1991} We have traced back the
differences with our results to the difference in the PPM used for the
ZrO$_2$, and to the inconsistence in the level of approximation for the
screened interaction $W$.  Our results combined with the DFT band
offsets available from the literature for different interfacial bonding
models provide values in a good agreement with the experiment.

\acknowledgments

M.G. thanks Dr.~Andrea Marini for his precious support for the use of the
\textsc{yambo} code. The authors acknowledge useful discussion with
Dr.~Matteo Giantomassi, Dr.~Patrick Rinke, Dr.~Alberto Zobelli, Dr.~Geoffrey
Pourtois and Prof. Michel~Houssa.  This work was supported by the EU's Sixth
and Seventh Framework Programs through the Nanoquanta Network of Excellence
(NMP4-CT-2004-50019),  the ETSF I3 e-Infrastructure project (Grant Agreement
211956), and the project FRFC N$^\circ$. 2.4502.05.  G.M.R. acknowledges the
FNRS of Belgium for financial support.

\bibliographystyle{apsrev}

\begin{thebibliography}{99}
\bibitem[{\citenamefont{Note3}(2009)\citenamefont{Note3}}]{note3}
\bibinfo{note}{For the calculation without the PPM, the inverse dielectric
matrix is computed explicitly for 100 frequencies ranging from 1 to 230 eV,
and the integration in Eq.~(\ref{eq:scrclb}) performed numerically.}
\end{thebibliography}

\end{document}